\documentclass[12pt]{article}
\def\be{\begin{equation}}
\def\ee{\end{equation}}
\def\bea{\begin{eqnarray}}
\def\eea{\end{eqnarray}}

\usepackage{graphicx}% Include figure files
\newcommand{\bi}{\bibitem}
\newcommand{\nn}{\nonumber}

\catcode`\@=11
\def\lsim{\mathrel{\mathpalette\@versim<}}
\def\gsim{\mathrel{\mathpalette\@versim>}}
\def\@versim#1#2{\vcenter{\offinterlineskip
\ialign{$\m@th#1\hfil##\hfil$\crcr#2\crcr\sim\crcr } }}
\catcode`\@=12
\usepackage{axodraw}

\catcode`@=12
\begin{document}
\thispagestyle{empty}

\begin{flushright}
UCRHEP-T409\\
KANAZAWA-06-01\\
July 2006\
\end{flushright}

\vspace{0.5in}

\begin{center}
{\LARGE \bf Cold Dark Matter, Radiative Neutrino Mass,\\
$\mu \to e \gamma$, and Neutrinoless Double Beta Decay\\}

\vspace{1.0in}
{\bf Jisuke Kubo$^a$, Ernest Ma$^b$, and Daijiro Suematsu$^a$\\}

\vspace{0.2in}
{\sl $^a$ Institute for Theoretical Physics, Kanazawa University, 
920-1192 Kanazawa, Japan\\}

\vspace{0.1in}
{\sl $^b$ Physics Department, University of California, Riverside,
California 92521, USA\\}

\end{center}

\vspace{1.0in}

\begin{abstract}\
Two of the most important and pressing questions in cosmology and particle
physics are: (1) What is the nature of cold dark matter? and (2) Will
near-future experiments on neutrinoless double beta decay be able to
ascertain that the neutrino is a Majorana particle, i.e. its own
antiparticle?  We show that these two seemingly unrelated issues are
intimately connected if neutrinos acquire mass only because of
their interactions with dark matter.
\end{abstract}

\newpage

\baselineskip 24pt

The existence of cold dark matter in the Universe is now well accepted
\cite{cdm}. From the viewpoint of particle physics, it should
consist of a weakly interacting yet-to-be-discovered neutral stable
fermion or boson.  A prime candidate is the lightest supersymmetric
particle (LSP) in the minimal supersymmetric standard model (MSSM).
More generally, we need only an exactly conserved $Z_2$ symmetry
\cite{m05,bhr06} and some new particles which are odd under it,
keeping all known particles even. In the MSSM, this $Z_2$ symmetry
is called $R$ parity, and the new particles are the squarks, sleptons,
gauginos, and higgsinos.

Consider now the interactions of the neutrino with particles in this
new class.  To realize the well-known dimension-five operator for Majorana
neutrino mass \cite{w79},
\begin{equation}
{\cal L}_{eff} = {f_{\alpha \beta} \over \Lambda} (\nu_\alpha \phi^0 -
l_\alpha \phi^+) (\nu_\beta \phi^0 - l_\beta \phi^+) + H.c.,
\end{equation}
where $(\nu_\alpha,l_\alpha)$ and $\Phi=(\phi^+,\phi^0)$ are the usual
lepton and Higgs
doublets of the standard model (SM), it is clear that the new
particles must form a loop with four external legs given by $\nu_\alpha
\nu_\beta \phi^0 \phi^0$.
There are generically three such one-loop diagrams
\cite{m98}.  In the MSSM, this does not happen because this operator
also requires lepton number to change by two units.  However, if a
neutral singlet superfield $N$ is added, then Figure 1 is generated
as a radiative contribution to the neutrino mass.
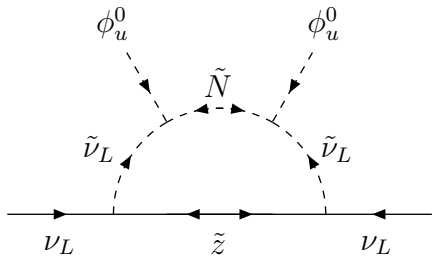
\begin{figure}[htb]
\begin{center}\begin{picture}(300,110)(0,45)
\ArrowLine(70,50)(110,50)
\ArrowLine(130,50)(190,50)
\ArrowLine(170,50)(110,50)
\ArrowLine(230,50)(190,50)
\Text(90,35)[b]{$\nu_L$}
\Text(210,35)[b]{$\nu_L$}
\Text(150,35)[b]{$\tilde z$}
\Text(150,100)[]{$\tilde N$}
\Text(105,70)[b]{$\tilde \nu_L$}
\Text(195,70)[b]{$\tilde \nu_L$}
\Text(110,116)[b]{$\phi^0_u$}
\Text(190,116)[b]{$\phi^0_u$}
\DashArrowLine(115,111)(130,85){3}
\DashArrowLine(185,111)(170,85){3}
\DashArrowArcn(150,50)(40,180,120){3}
\DashArrowArcn(150,50)(40,100,60){3}
%\Vertex(150,90){3}\Vertex(150,50){3}
\DashArrowArc(150,50)(40,0,60){3}
\DashArrowArc(150,50)(40,80,120){3}
%\Vertex(190,50){3}\Vertex(150,90){3}
\end{picture}\end{center}
\caption[]{One-loop radiative neutrino mass in supersymmetry.}
\end{figure}

We note that the particles in the loop all have odd $R$ parity.
Of course, in the context of supersymmetry, this also implies that
$\nu$ couples to $N$ directly through $\phi_u^0$, so that a Dirac mass
already appears, and together with the heavy Majorana mass of $N$, the
famous seesaw mechanism \cite{seesaw} allows $\nu$ to obtain a tree-level
Majorana mass.  That is why Figure 1 is always negligible in the
MSSM with the addition of $N$.  However, in a more general framework,
the new particles in the loop may be the only source of neutrino mass
\cite{kst03,m06}, and in that case there will be interesting
phenomenological implications on lepton flavor transitions and
neutrinoless double beta decay, as shown below.

Basically, the argument goes as follows.  In order for the new particles
in the loop to be identified as cold dark matter with the correct
value of their relic density in the Universe at present, their
interactions with neutrinos and charged leptons must not be too weak.
On the other
hand, they are also responsible for the masses of neutrinos and their
observed mixing in neutrino oscillations.  This implies necessarily
flavor changing transitions such as $\mu \to e \gamma$.
In order to suppress the latter, the parameter space of
neutrino masses is limited, thereby
enforcing a lower bound on neutrinoless double beta decay.

Of the three generic one-loop diagrams giving rise to a radiative neutrino
mass, the simplest in terms of new particle content is given in
Ref.~\cite{m06}.  The standard model is extended by adding three
neutral singlet fermions $N_i$ and a second scalar doublet $(\eta^+,\eta^0)$,
which are odd under an exactly conserved $Z_2$ symmetry, keeping all SM
particles even.  In that case, the analog of Figure 1 is Figure 2, as
depicted below.
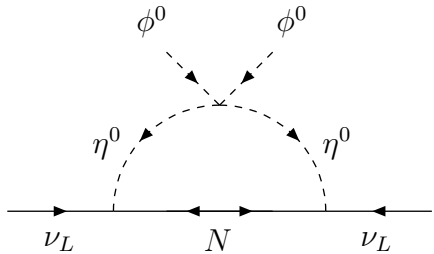
\begin{figure}[htb]
\begin{center}\begin{picture}(300,110)(0,45)
\ArrowLine(70,50)(110,50)
\ArrowLine(130,50)(190,50)
\ArrowLine(170,50)(110,50)
\ArrowLine(230,50)(190,50)
\Text(90,35)[b]{$\nu_L$}
\Text(210,35)[b]{$\nu_L$}
\Text(150,35)[b]{$N$}
\Text(108,70)[b]{$\eta^0$}
\Text(195,70)[b]{$\eta^0$}
\Text(125,116)[b]{$\phi^0$}
\Text(178,116)[b]{$\phi^0$}
\DashArrowLine(130,110)(150,90){3}
\DashArrowLine(170,110)(150,90){3}
\DashArrowArc(150,50)(40,90,180){3}
%\Vertex(150,90){3}\Vertex(150,50){3}
\DashArrowArcn(150,50)(40,90,0){3}
%\Vertex(190,50){3}\Vertex(150,90){3}
\end{picture}\end{center}
\caption[]{One-loop radiative neutrino mass in the model of Ref.~\cite{m06}.}
\end{figure}

We note again that the particles in the loop are all odd, and that lepton
number changes by two units as in Figure 1.  The $Z_2$ invariant
Higgs potential is given by
\bea
V &=& m_1^2 \Phi^{\dag}\Phi+m_2^2\eta^{\dag}\eta
+\frac{1}{2}\lambda_1 (\Phi^{\dag}\Phi)^2
+\frac{1}{2}\lambda_2 (\eta^{\dag}\eta)^2+
\lambda_3 (\Phi^{\dag}\Phi)(\eta^{\dag}\eta)\nn\\
& &+~\lambda_4  (\Phi^{\dag}\eta)(\eta^{\dag}\Phi)+
\frac{1}{2}\lambda_5[(\Phi^{\dag}\eta)^2+H.c.],
\label{pot}
\eea
with $\langle \phi^0 \rangle = v$ and $\langle \eta^0 \rangle = 0$.  Let
us choose the bases where $N_i$
and $l_\alpha,l^c_\alpha$ are diagonal, and consider the interactions of
$(\nu_\alpha,l_\alpha)$ with $N_i$ and $(\eta^+,\eta^0)$, i.e.
\bea
{\cal L}_N &=&h_{\alpha i}(\nu_\alpha\eta^0-l_\alpha \eta^+)N_i
+ H.c.
\label{yukawa}
\eea
Since $\langle \eta^0 \rangle = 0$ is required to preserve the
exact $Z_2$ symmetry, there are no Dirac masses linking $\nu_\alpha$ with
$N_i$.  In other words, even though $N_i$ have heavy Majorana masses, the
canonical seesaw mechanism is not operative.  Further, the lightest among
the new particles will be stable and becomes an excellent
candidate for the cold dark matter of the Universe.  We see thus that
{\it neutrinos acquire mass here only because of their
interactions with dark matter}.

We note that Figure 2 depends on the existence of the $\lambda_5$ coupling of
Eq.~(\ref{pot}).  If it were zero, we could assign the exactly conserved
additive lepton number
$-1$ to $(\eta^+,\eta^0)$ and 0 to $N_i$, in which case the neutrinos would
stay massless.  This means that it is natural for $\lambda_5$ to be very
small,
which we will assume from here on.  Without loss of generality, $\lambda_5$
may also be chosen to be real.  We now define $\eta^0 = (\eta_R + i \eta_I)
/\sqrt 2$ and obtain $m^2_R-m^2_I =2 \lambda_5 v^2$, where $m_R (m_I)$ is
the mass of $\eta_R (\eta_I)$.  Using $m_0^2=(m_R^2+m_I^2)/2$, the
radiative neutrino mass matrix is then given by \cite{m06}
\bea
({\cal M}_\nu)_{\alpha \beta} = \sum_i {h_{\alpha i} h_{\beta i}
I(M_i^2/m_0^2) \over M_i},
\label{mnu}
\eea
where
\bea
I(x) = {\lambda_5 v^2 \over 8 \pi^2}
\left(
{x \over 1 -
x} \right) \left[ 1 + {x \ln x \over 1-x}
\right].
\label{func-i}
\eea

Assuming that atmospheric neutrino mixing is maximal, the neutrino mixing
matrix $U$ which diagonalizes ${\cal M}_\nu$ can be written as
$U ={\hat U} P$,
where ${\hat U}$ is approximately given by
\bea
{\hat U} &\simeq &
\pmatrix{c_{12} & s_{12} & s_{13}e^{-i\delta} \cr 
-s_{12}/\sqrt 2 + c_{12} s_{13} e^{i \delta}/\sqrt 2 & 
c_{12}/\sqrt 2 + s_{12} s_{13} e^{i \delta}/\sqrt 2 & 
-1/\sqrt 2 \cr 
-s_{12}/\sqrt 2 - c_{12} s_{13} e^{i \delta}/\sqrt 2 
 & c_{12}/\sqrt 2 - s_{12} s_{13} e^{i \delta}/\sqrt 2 & 1/\sqrt 2},
\label{u}\\
P & =& \pmatrix{ e^{i \alpha_1/2} & 
0 & 0\cr 
0 &  e^{i \alpha_2/2}& 0 \cr
 0 &0 & 1},
\eea
and $c_{12}=\cos\theta_{12}$, $s_{12} = \sin \theta_{12}$,
$s_{13}=\sin\theta_{13}$, with $\tan^2 \theta_{12} \simeq 0.45$,
and $s_{13} \lsim 0.2$.

We now assume the lightest $N$ to be dark matter.  Call it $N_k$.
We need to calculate its relic density as a function of its interaction 
strengths $h_{\alpha k}$, its mass $M_k$, and the masses of $\eta^\pm$,
$\eta_R$, and $\eta_I$, which we take for simplicity to be all given by
$m_0$, with $m_0 > M_k$.  Our goal is to obtain the observed dark-matter
relic density of $\Omega_d h^2 \simeq 0.12$ \cite{wmap,sdss}.

The thermally averaged cross section for the annihilation of two $N_k$'s
into two leptons is computed by expanding the corresponding relativistic
cross section $\sigma$ in powers of their relative velocity and keeping
only the first two terms.  Using the result of Ref.~\cite{griest1}, and
recognizing that lepton masses are very small, we have
\bea
\langle \sigma v \rangle = a + b_k v^2+\cdots,~~~~~
a =0, ~~~ b_k = \frac{y_k^4 r_k^2 (1-2 r_k+2 r_k^2)}{24 \pi M_k^2},
\label{bandr}
\eea
where
\bea
  r_k =  M_k^2/(m_0^2+M_k^2),~~~
  y_k^4=\sum_{\alpha\beta}|h_{\alpha k}h_{\beta k}^*|^2.
  \label{randy}
\eea
Following Ref.~\cite{griest2}, the relic density of $N_k$ is then given by
$\Omega_d h^2 = Y_\infty s_0 M_k h^2/\rho_c$,
where $Y_\infty$ is the asymptotic value of the ratio $n_{N_k}/s$, with
$Y_\infty^{-1} = 0.264 g_*^{1/2} M_{\rm Pl} M_k (3 b_k/x_f^2)$,
$s_0=2970/\mbox{cm}^3$ is the entropy density at present,
$\rho_c=3 H^2/8 \pi G=
1.05 \times 10^{-5}h^2 ~\mbox{GeV}/\mbox{cm}^3$ is the critical density,
$h$ is the dimensionless Hubble parameter,
 $M_{\rm Pl}=1.22\times 10^{19}~ \mbox{GeV}$ is the  Planck energy,
and $g_*$ is the number of effectively massless degrees of freedom
at the freeze-out temperature.
Further, $x_f$ is the ratio $ M_k/T$ at the
freeze-out temperature and is given by
\bea
x_f &=&\ln \frac{0.0764 M_{\rm Pl}(6 b_k/x_f) c (2+c) M_k}
{(g_* x_f)^{1/2} }.
\label{xf}
\eea
Using $g_*^{1/2}=10$ and $c=1/2$ as in Ref.~\cite{griest2}, we obtain
\bea
\left[ \frac{ M_k}{\mbox{GeV}} \right] &=&5.86\times 10^{-8}~
 x^{-1/2}_f~e^{x_f}~
\left[\frac{\Omega_d h^2}{0.12}\right],
\label{mn}\\
\left[ \frac{b_k }{\mbox{(GeV)}^{-2}} \right] &=& 2.44 \times 10^{-11}~
 x_f^2~\left[\frac{0.12}{\Omega_d h^2}\right],
\label{meta}
\eea
where $b_k$ and $y_k$  are given in Eqs.~(\ref{bandr}) and (\ref{randy}).
Since $b_k/y_k^4$ is a
function of $ M_k$ and $m_0$, Eqs.~(\ref{mn}) and (\ref{meta})
allow us to calculate $ M_k$ and $m_0$ in units of GeV for a given set
of $y_k^2, x_f$ and $\Omega_d h^2$.
\begin{figure}[htb]
\includegraphics*[width=0.8\textwidth]{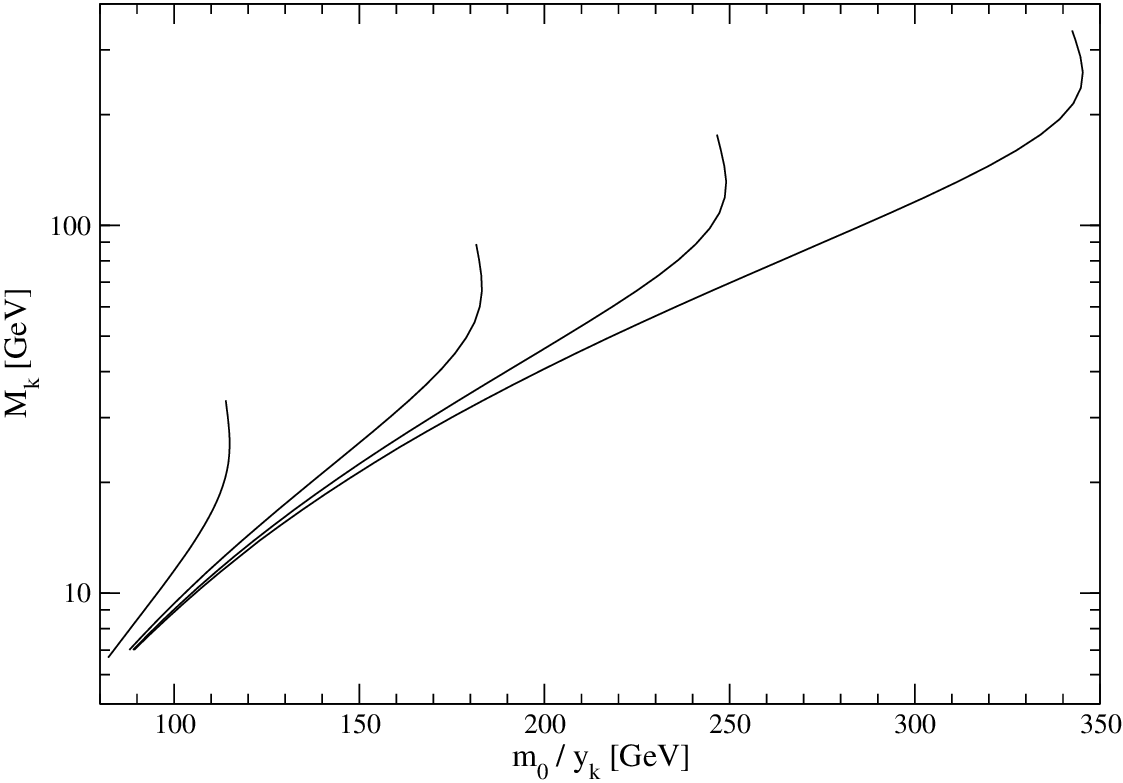}
\caption{\label{mk-m0-y}\footnotesize
$M_k$ versus $m_0/y_k$ for $y_k=0.3, 0.5, 0.7, 1.0$ (left to right)
for $\Omega_d h^2=0.12$,
where $y_k$ is defined in Eq.~(\ref{randy}).}
\end{figure}

In Figure~\ref{mk-m0-y}, we plot $M_k$ versus $m_0/y_k$
for $y_k =0.3, 0.5, 0.7, 1.0$.  As we can see from the figure, the dark
matter constraint requires that $M_k$ increases as $m_0$ increases and
for each value of $m_0$, it may be as large as $m_0$.  We also see that
$m_0/y_k$ cannot exceed $350$ GeV or so in the perturbative regime
$y_k \lsim 1$.  This is a very strong constraint,
because $m_0/y_k$ sets the scale also for lepton flavor transitions such as
$\mu \to e \gamma$ and the experimental upper bound of its branching
fraction cannot be satisfied, unless some cancellation mechanism
is at work.

The branching fraction of $\mu \to e \gamma$ is given in this model by
\cite{mr01}
\bea
B(\mu\to e\gamma)
=\frac{3\alpha}{64\pi (G_F m_0^2)^2} C^4
\simeq \left(\frac{30~\mbox{GeV}}{m_0/C}\right)^4,
 \label{mutoe}
\eea
where
\bea
C^2 =  \left| \sum_i h_{\mu i} h^*_{e i}  
F_2(M_i^2/m_0^2) \right|,
\label{mueg}
\eea
and
\bea
F_2(x) = {1-6x+3x^2+2x^3-6x^2\ln x \over 6(1-x)^4}.
\label{func-f2}
\eea
Since $M_k < m_0$ should be satisfied for $N_k$ dark matter, the function
$F_2(x_k)$ can vary only between $1/12 (x_k=1)$ and $1/6 (x_k=0)$.
To suppress the branching fraction $B(\mu\to e\gamma)$ which is inversely
proportional to the fourth power of $m_0$, we need a large value of $m_0$.
On the other hand, the observed dark matter relic density
$\Omega_d h^2\simeq0.12$ requires $m_0$ to be below 350 GeV for
$y_k = (\sum_{\alpha\beta}|h_{\alpha k}h_{\beta k}^*|^2)^{1/4} \lsim 1$.
This means that if $|\sum_i h_{\mu i} h^*_{e i}|$ appearing in
Eq.~(\ref{mueg}) is also of order 1, then $B(\mu\to e\gamma) \gsim
5\times10^{-7}$, which is several orders of magnitude above the
experimental upper bound of $1.2\times 10^{-11}$.

To satisfy the $\mu \to e \gamma$ constraint, we consider the possibility
that $M_{1,2,3}$ are nearly degenerate.  In that limit,
\bea
({\cal M}_\nu)_{\alpha \beta} =
{I(M^2/m_0^2) \over M} \sum_i h_{\alpha i}
h_{\beta i} = {\hat U}^*
 \pmatrix{m_1e^{-i\alpha_1} & 0 & 0 \cr 0 & m_2
 e^{-i\alpha_2} & 0 \cr 0 & 0 & m_3}
  {\hat U}^\dag,
\eea
where ${\hat U}$ is given by Eq.~(\ref{u}).
A simple solution is then
\bea
h_{\alpha i} =
\left[ {M m_i \over I(M^2/m_0^2)}
 \right]^{1/2}~e^{-i \alpha_i/2}~ {\hat U}_{\alpha i}^*
 ~\mbox{with}~\alpha_3=0.
\eea
Then we obtain
\bea
C^2 &=& {F_2(M^2/m_0^2) M \over I(M^2/m_0^2)}
\left|\sum_i {\hat U}_{\mu i}
{\hat U}^*_{e i} m_i \right| \nonumber \\
&=& {F_2(M^2/m_0^2) M \over I(M^2/m_0^2)}
\left| {s_{12} c_{12} \over \sqrt 2} (m_2-m_1)
 + {s_{13} e^{-i\delta} \over
\sqrt 2} (c_{12}^2 m_1 + s_{12}^2 m_2 - m_3) \right|.
\eea
Thus the suppression of $C^2$ is possible because $m_2-m_1$ is related
to $\Delta m^2_{sol}$ and
$c_{12}^2 m_1 + s_{12}^2 m_2 - m_3$ is related
to $\Delta m^2_{atm}$ in neutrino oscillations \cite{mr01,x01}.
\begin{figure}[htb]
\includegraphics*[width=0.8\textwidth]{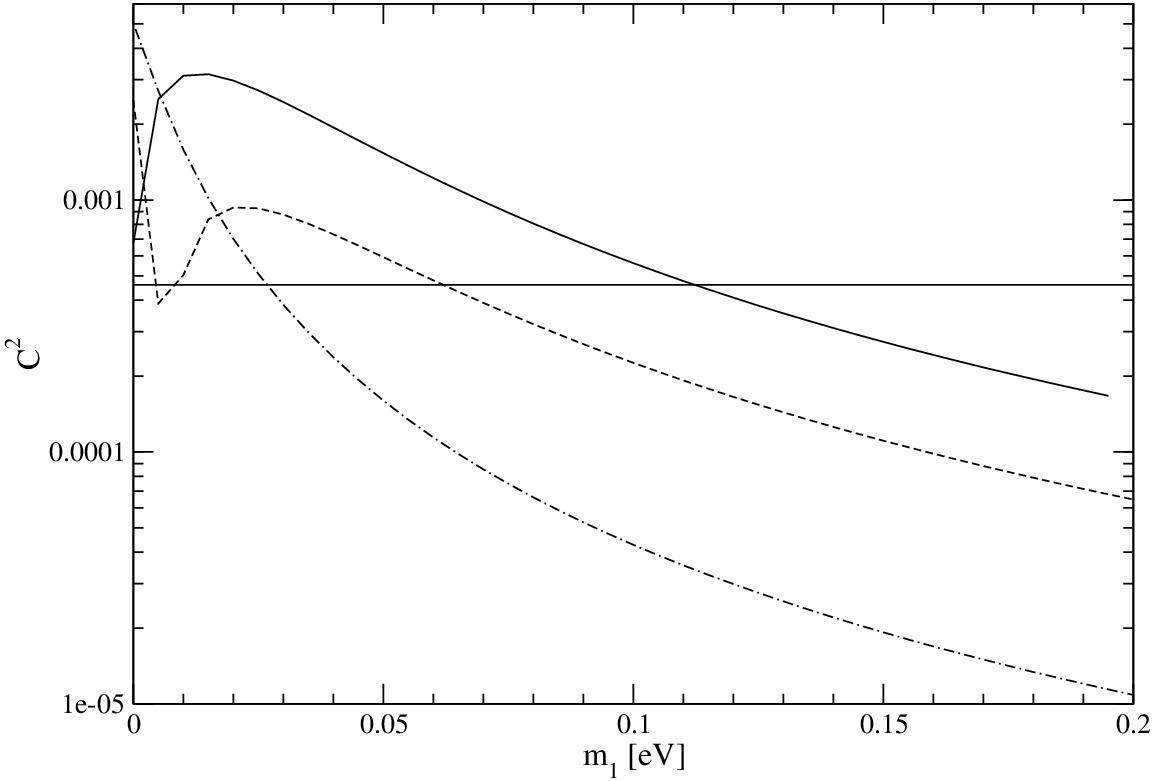}
\caption{\label{c-m1}\footnotesize
$C^2$ versus $m_1$ in the case of normal ordering for $s_{13}=
0.1~(\mbox{solid}),
0.05~(\mbox{dash}),
0.01~(\mbox{dot-dash})$, where
$C^2=4.6\times 10^{-4}$ (horizontal line)
corresponds
to the experimental
upper bound $B(\mu\to e\gamma)=1.2\times 10^{-11}$.
}
\end{figure}

Let us assume $\delta = 0$ and consider the normal ordering of neutrino
masses, i.e. $m_3$ is the largest mass.  We then set
$h_3 = \left(
\sum_\alpha |h_{\alpha 3}|^2 \right)^{1/2} = 1$ which is equivalent to
having $M m_3/I(M^2/m_0^2) = 1$. Hence
\bea
C^2 \simeq \left( {0.067 \over m_3} \right) |c_{12}(s_{12}-s_{13}c_{12})
(m_2-m_1) - s_{13}(m_3-m_2)| < 4.6 \times 10^{-4},
\eea
where $F_2 = 0.0948$ (corresponding to $m_0=345$ GeV and $M=290$ GeV).
Using $\Delta m^2_{21} = \Delta m^2_{sol} =
7.9 \times 10^{-5}$ eV$^2$ and $\Delta m_{32}^2 =
\Delta m^2_{atm} = 2.3 \times 10^{-3}$ eV$^2$,
we plot $C^2$ versus $m_1$ in Figure~\ref{c-m1} for $s_{13}
=0.1, 0.05, 0.01$.  The horizontal line
is the experimental bound
$C^2=4.6\times 10^{-4}$ corresponding
to $B(\mu\to e\gamma)=1.2\times 10^{-11}$.
We find that for $s_{13} \gsim 0.26$, this
constraint cannot be satisfied.
For $s_{13}$ less than its experimental bound of $0.2$, there is a lower
bound on $m_1$ according to the approximate empirical formula
\bea
\left[ {m_1 \over {\rm eV}} \right] \gsim 0.02 + 1.4 |s_{13}-0.02|
- 2.9 |s_{13}-0.02|^2,
\eea
except for a tiny region near $m_1=0$ and $s_{13} = 0.09$, and a small
region near $m_1=0.01$ eV and $s_{13} = 0.04$.  In Figure~\ref{c-m1}, we
can see that the plot for $s_{13}=0.1$ is getting close to the first
region from its dip at $m_1=0$, and that the plot for $s_{13}=0.05$ has
a small allowed range near $m_1=0.01$ eV.
\begin{figure}[htb]
\includegraphics*[width=0.8\textwidth]{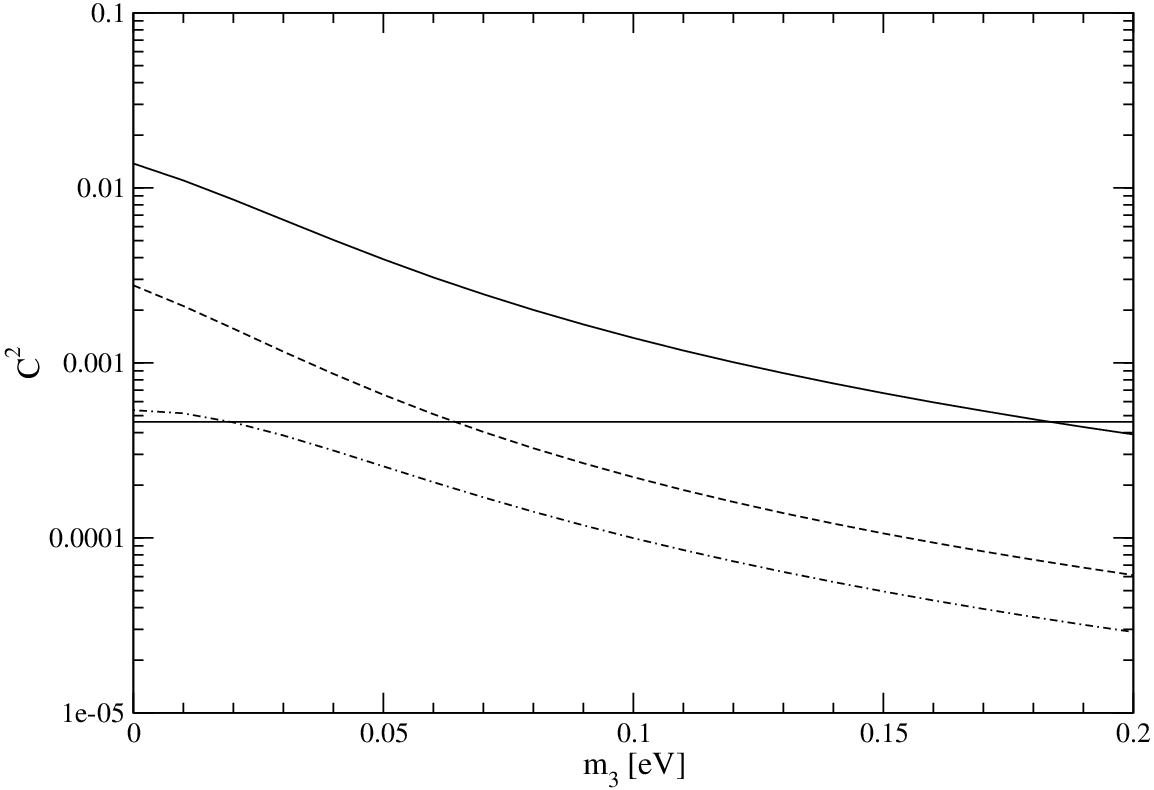}
\caption{\label{c-m3}\footnotesize
$C^2$ versus $m_3$ in the case of inverted ordering for
$s_{13}=0.2~(\mbox{solid}),-0.05~(\mbox{dash}),
0.0~(\mbox{dot-dash})$.
}
\end{figure}

In the case of inverted ordering, i.e. $m_2$ is the largest mass, we
are already guaranteed that $m_{1,2} > \sqrt{\Delta m^2_{32}} \simeq
0.048$ eV.  For completeness, we set $h_2=1$ and plot $C^2$ versus $m_3$
in Figure~\ref{c-m3} for $s_{13}=0.2, -0.05, 0.0$.
Here, for $s_{13} \gsim 0.24$ and $s_{13} \lsim -0.27$,
the experimental constraint cannot be satisfied.  In other words,
the constraint on $\theta_{13}$ from $\mu \to e \gamma$ coincides
roughly also with that from neutrino oscillations.

For the simple solution of Eq.~(16), 
as we can see from Figure 4 and 5, all the neutrino masses may be assumed
to be degenerate to satisfy the $\mu \to e\gamma$ constraint; hence
the effective mass $\langle m_{ee} \rangle$ in neutrinoless double beta
decay is approximately given by
\bea
\langle m_{ee} \rangle \simeq m_1 |
0.572 + 0.428 \cos(\alpha_1-\alpha_2)|^{1/2}.
\eea
We also allow the heavy $N_i$ masses $M_{1,2,3}$ to be
slightly different, so that our approximation that only one of them is
the candidate for dark matter remains valid.  Note that $\Delta M/M$
only needs to be of order $10^{-3}$ for Eq.~(18) to be valid.
In analogy to $\mu \to e \gamma$, there are also dark-matter contributions
to $\tau \to \mu \gamma$, $\tau \to e \gamma$, and the anomalous magnetic
moment of the muon.  However, they are at least one or more orders of
magnitude below the present experimental bounds.

In conclusion, we have shown how cold dark matter and neutrinoless double
beta decay may be connected if neutrinos acquire mass only because of
their interactions with dark matter.  We repeat the basic argument
presented earlier.  The existence of dark matter requires a class of
new particles which are odd with respect to an exactly conserved $Z_2$
symmetry.  Their interactions with neutrinos and charged leptons must
not be too weak to be identified as cold dark matter with the correct
value of their relic density in the Universe at present.  On the other
hand, they are also responsible for the masses of neutrinos and their
observed mixing in neutrino oscillations.  This implies necessarily
flavor changing transitions such as $\mu \to e \gamma$.  In order to
suppress the latter, the parameter space of neutrino masses is limited,
thereby enforcing a lower bound on neutrinoless double beta decay.
For $N$ as dark matter, this is typically of order 0.05 eV, even though
much lower values are still allowed from accidental cancellations.  More
importantly, this connection between cold dark matter and neutrinoless
double beta decay can be tested in the near future at the Large Hadron
Collider and complemented by a host of experiments on neutrino oscillations
and neutrinoless double beta decay already under way and being planned.

%\vspace{0.5cm}
This work was supported in part by the U.~S.~Department of Energy under
Grant No. DE-FG03-94ER40837.  We thank Haruhiko Terao for valuable
discussions.  EM also thanks the Institute for Theoretical
Physics, Kanazawa University for hospitality during a recent visit.

%\newpage

\bibliographystyle{unsrt}

\end{document}